# Diverse electronic and magnetic properties of CrS$_2$ enabling novel strain-controlled 2D lateral heterostructure spintronic devices

Kaiyun Chen[1], Junkai Deng[1, *], Yuan Yan[1,2], Qian Shi[1], Tieyan Chang[1], Xiangdong Ding[1], Jun Sun[1], Sen Yang[1, *] and Jefferson Zhe Liu[2, *]

[1] MOE Key Laboratory for Nonequilibrium Synthesis and Modulation of Condensed Matter, State Key Laboratory for Mechanical Behavior of Materials, Xi'an Jiaotong University, Xi'an 710049, China

[2] Department of Mechanical Engineering, The University of Melbourne, Parkville, VIC 3010, Australia

*junkai.deng@mail.xjtu.edu.cn; *yangsen@mail.xjtu.edu.cn; *zhe.liu@unimelb.edu.au

*Corresponding authors

## Abstract

Lateral heterostructures of two-dimensional (2D) materials, integrating different phases or materials into a single piece of nanosheet, have attracted intensive research interests in the past few years for high-performance electronic and optoelectronic devices. It also holds promises to significantly improve the performance and enable new functions of spintronic devices. It is imperative to have a 2D material possessing diverse electronic and magnetic properties that are required in spintronics. In this work, using density functional theory calculations, we surveyed all IV, V and VI group transition metal dichalcogenides (TMDs) and discovered that CrS$_2$ has the most diverse electronic and magnetic properties: antiferromagnetic (AFM) metallic 1T phase, nonmagnetic (NM) semiconductor 2H phase, and ferromagnetic (FM) semiconductor 1T' phase with a Curie temperature of ~1000 K. More interestingly, we found that a tensile or compressive strain could turn 1T' phase into a spin-up or spin-down half metal. Such a unique feature enables designing strain-controlled spintronic devices using a single piece of CrS$_2$ crystal with improved energy efficiency, which remains a challenge in miniaturization of spintronic devices. In-depth analysis attributed the unique strain tunability to the interplay between strain-induced lattice deformation and different spatial orientation of the spin-up/spin-down electronic orbitals. A prototypical design of a simple spin-valve logic device operated by strain is also presented.

# Introduction

Spintronics utilizes two intrinsic properties of electrons, spin and charge, to represent information compared with conventional charge-based electronics devices.[1-4] It is considered as one of the most important emerging research areas with an immense potential to provide high speed, low power and high density logic and memory electronic devices. In the past few decades, many novel spintronic phenomena were discovered including giant magnetoresistance (GMR), tunneling magnetoresistance (TMR), topological insulators, and quantum spin Hall effect.[5-8] Magnetoresistive random access memory (MRAM), one type of spintronics devices, is the most promising candidate for high density storage, ultrafast data processing and nonvolatility novel information devices.[9,10] The basic structure of MRAM is a spin valve – a three-layer sandwich structure consisted of two ferromagnetic (FM) layers on two sides and one non-FM interlayer in middle. The parallel or antiparallel magnetization of two FM layers generates a low or high resistance, designated as '1' or '0' bit. The writing operation is done via applying a magnetic field or using spin torque transfer to control magnetism of the two FM layers.[11,12] These methods, however, require a high current density flowing in devices and lead to high energy consumption. Some other problems include high temperature reducing equipment reliability, energy waste, and low efficiency. Lower energy consumption is a challenge in spin valve miniaturization.[13]

There have been many efforts to utilize electric field to replace magnetic field or spin torque transfer for a low energy consumption spin valve. Some previous works showed that coercivity and magnetization could be changed under an electric field, e.g., voltage-controlled magnetoresistance realized in CoFeB/MgO/CoFeB and Fe/MgO/FeCo.[14,15] Unfortunately, a bias magnetic field is still required and the voltage applied to the devices is close to the breakdown voltage. Another option of utilizing electric field is multiferroic materials. The interlayer of a spin value should be thin enough. But ferroelectricity or multiferroicity is hard to maintain in films with a thickness of just a few nanometers. In addition, the operation temperature is not practical in real applications.[16,17] To overcome these problems, a new fabrication method is to deposit a spin valve made of magnetostriction materials on a ferroelectric (piezoelectric) substrate[18,19], *e.g.,* a MgO-based spin valve Ta(5 nm)/Ru(5 nm)/IrMn(8 nm)/CoFe(2.3 nm)/Ru(0.85 nm)/CoFeB(2.6 nm)/MgO(2.3 nm)/CoFeB(2.6 nm)/Ta(5 nm)/Ru(7 nm) deposited on $Pb(Mg_{1/3}Nb_{2/3})_{0.7}Ti_{0.3}O_3$ substrate. These devices could be regarded as a strain-mediated spin valve. The working mechanism is to transfer a piezoelectric strain generated from the substrate to magnetostriction layers, and the subsequently generated magnetic field can change the magnetization of the two FM layers. Such strain-controlled spin valve has low energy consumption.[20,21] But these devices have complex structures (*e.g.*, 11 layers of different materials).

Two-dimensional material heterostructures, including multi-layer stacking or lateral interfacing, provide great opportunities to fabricate high-performance electronics and optoelectronic devices. A lateral heterostructure combines different materials or phases into a single piece of nanosheet. It has an atomically sharp interface and junction region compared with a multilayer stacked heterostructure.[22] An ultrathin transistor consisting of a metallic 1T phase and semiconductor 2H phase on single $MoS_2$ nanosheet was fabricated and showed a low contact resistance.[23] Besides, in a lateral heterostructure, local regions can be regulated under some external stimuli to realized different properties in one device, enabling novel functions.[24-27] The 2D material lateral heterostructure could be a promising solution for low energy consumption high-performance spin valves.[28] As such, a 2D material that possesses all the required properties in spin valve is indispensable. Transition metal dichalcogenide (TMD) is the most attractive material group. It has rich chemical diversity as well as multiple phases (2H, 1T, 1T', and 1T''), leading to diverse properties including direct band gap, strong spin−orbit coupling, catalytic and robust yet flexible mechanical property.[29] Besides, some physical properties are sensitive to external stimuli, such as strain engineering band gap and magnetism.[30,31]

In this paper, using density function theory (DFT) calculation, we systematically surveyed all IV, V and VI group TMDs and discovered that $CrS_2$ has the most diverse electronic and magnetic properties that are required in spintronics, including antiferromagnetic (AFM) metallic 1T phase, nonmagnetic (NM) semiconductor 2H phase, and ferromagnetic (FM) semiconductor 1T' phase. The most interesting observation is that a tensile or compressive strain can turn 1T' $CrS_2$ into a spin-up or spin-down half metal, which is not observed in other 2D materials before. Half metal is highly desirable in spintronics as it ensures high purity spin current. The strain tunable half metallicity of $CrS_2$ is a unique feature that allows strain-controlled 2D lateral heterostructure spintronic devices (such as spin valve) in a single piece of $CrS_2$ nanosheet, which could have the advantages of low energy consumption as well as a relatively simple device structure. In-depth analysis was carried out to understand the special strain tunability. As a demo, a prototype design of strain-controlled spin valve logic device is also presented in the end of this work.

## Results and discussion

### Electronic and magnetic properties of 2H, 1T and 1T' phases of TMDs

Figure 1a depicts the crystal structure of 2H, 1T and 1T' phase of TMDs ($MX_2$, M as transition metal element and X as S, Se or Te). The primitive unit cell of 2H phase is a 120° rhombus with $P6_3/mmc$ space group (orange region in Figure 1a). 1T is a high symmetry phase with $P\bar{3}m1$ space group and a 120° rhombus primitive cell. It can be transformed from the 2H phase via a collectively sliding of the top layer S atoms to the hexagon centers. The 1T' can be regarded as a distorted structure from the 1T phase, arising from the Pierels instability.[32] Its primitive cell is a rectangle unit, corresponding to a ($1\times\sqrt{3}\times1$) supercell of the 1T phase (green rectangle in Figure 1a). Compared with the 1T phase, two adjacent lines of metal atoms in the high symmetry *y*-direction move toward each other to form a zig-zag M-M dimer chain (the dashed lines in Figure 1a). Hence the distance between the adjacent metal atoms lines (along *y*-direction) are not equal in 1T' phase: $d_1/d_2 < 1$, different from a 1T phase having $d_1/d_2 = 1$. For the sake of a consistent comparison, all our DFT calculations for the 2H, 1T and 1T' phases were based on the rectangle unit cell.

We carried out a comprehensive computational survey of 2H, 1T and 1T' phases of all possible group IV, V and VI element TMDs. The relative total energies of these three phases and their magnetic properties are shown in Figure S1 and S2 (supplementary information), respectively. A clear trend can be observed in Figure S1. For group IV TMD, 1T is the most stable phase and 1T' phase does not exist. As moving to the right of the periodic table, *i.e.*, M atoms changing to the V and VI groups, the 1T' phase appears and the 2H phase becomes the ground state for most TMDs. All three phases can exit in the group VI TMDs. In addition, as we move from group IV to group VI, more diverse magnetic properties can be observed in Figure S2. All compounds in group IV TMDs are non-magnetism (NM). Both ferromagnetism (FM) and anti-ferromagnetism (AFM) can be observed in group IV and VI TMDs, such as FM-1T-$VS_2$ and AFM-1T-$CrS_2$. Figure 1b summarizes all results including (mechanical) stable phases, electronic conductivity, and magnetism. The blue background highlights the cases with stable 2H and 1T phases and the orange background represents the cases with three stable phases. A clear boundary exists in the group V TMDs to separate the two states.

There are four different types of combination of electronic conductivity (metal vs. semiconductor) and magnetism (NM vs. AFM/FM). Among all TMDs compounds in Figure 1b, $CrS_2$ is a special one with the most diverse electronic/magnetic properties (in its three different phases), *i.e.*, three out of the four possible combinations. The electronic band structures of the 2H- and 1T-$CrS_2$ are shown in Figure S3, and 1T' band structure is shown in Figure 2a. All these band structures were calculated using the Heyd-Scuseria-Ernzerh of (HSE06) hybrid functional. The band structures indicate that 2H-$CrS_2$ is nonmagnetic semiconductor (NM-SC) and 1T-$CrS_2$ is antiferromagnetic metal (AFM-M), which agree with previous studies.[33,34] It is worth noting that the $CrS_2$ 1T' phase not reported before is ferromagnetic semiconductor (FM-SC). Our DFT calculation showed that it had a total energy value higher than the 2H phase but lower than the 1T phase (Figure S1). The calculated phonon spectrum in Figure S4 indicates the mechanical stability of this new phase. Figure S5 and Table S1 summarize different magnetic states of 1T'-$CrS_2$. Our results conclude that FM semiconductor is the ground state of 1T'-$CrS_2$.

The 1T' FM-SC phase can be viewed as a distorted 1T AFM-M phase (Figure 1a). The observed lattice distortion, two adjacent rows of Cr atoms moving toward each other, should stabilize the FM state over the AFM state in 1T phase. The spin polarized charge density in Figure S6 clearly shows such transition. To gain some in-depth understanding, we compared the partial density of states (PDOS) results of Cr $d$ and S $p$ electrons for these two phases in Figure S7a and b, respectively. For 1T phase, the Cr $d$ orbital is spin polarized, making the main contribution (2.664 $\mu_B$ per Cr atom) to the total magnetism (Figure S7a), whereas S atoms have very weak magnetic moments 0.042 $\mu_B$. The magnetic interactions between two adjacent lines of Cr atoms should be the direct-exchange-interaction.[31,35] For 1T' phase, Figure 7b shows that the hybridization between Cr $d$ and S $p$ orbitals is stronger. The magnetic moment of S atoms significantly increases to $-0.326\mu_B$. The enhanced magnetic moments of S atoms and their opposite signs to those of Cr atoms suggest the super-exchange-interaction (Cr-S-Cr interaction) in 1T' phase. Thus, the lattice distortion triggers the super-exchange-interaction and thus stabilizes the FM state in 1T' phase.

The band structure around Fermi level of 1T'-CrS$_2$ is shown in Figure 2a. There is a small gap for spin-up electrons (0.26 eV along Y-S) and a large gap for spin-down electrons (1.92eV along Γ-S). Both spin-up and down band gaps are indirect. When the 1T AFM-M phase transforms to 1T' FM-SC phase, the energy gaps are opened for both spin-up and spin-down electrons. For many TMDs, the phase transition from 1T to 1T' phase often leads to a band gap opening. [32]

As a summary, Table 1 lists the crystal structure, magnetism, band gap, and relative total energy results of the three different CrS$_2$ phases, 2H (NM-SC), 1T (AFM-M), and 1T' (FM-SC). On top of such most diverse electronic/magnetic properties of CrS$_2$ among all TMDs, we also found that strain can be used to transit FM-S 1T' phase into spin-up and spin-down half metals. This result is very interesting in light of the promising potential in spintronics application.

**Strain controlled half-metal switching in 1T'-FM-CrS$_2$**

A uniaxial tensile ($\varepsilon_y$ = +6%) and compressive ($\varepsilon_y$ = −6%) strain along $y$-direction were applied on 1T'-CrS$_2$ to investigate the effects of strain on 1T'-CrS$_2$ electronic structure. Figure 2 shows the band structures of the strained 1T' phases using Heyd–Scuseria–Ernzerhof hybrid functional (HSE06). The left figure in Figure 2a shows that the compressive strain ($\varepsilon_y$ = −6%) closes the spin-down electron band gap from 1.92 eV to zero, but open the spin-up band gap from 0.26 eV to 0.62 eV. In contrast, the right figure in Figure 2a shows that the tensile strain ($\varepsilon_y$ = +6%) closes the spin-up band gap from 0.26 eV to zero but open the spin-down band gap from 1.92 eV to 2.62 eV. Clearly, a compressive and tensile strain can transform the 1T'-CrS$_2$ into spin-down and spin-up half metal, respectively, from a ferromagnetic semiconductor. For comparison, Figure S8 also shows the calculated band structures under 0%, +6% and −6% strain using the Perdew-Burke-Ernzerhof (PBE) functional. The same conclusion can be drawn.

To understand the transition, the spin polarized band structures under a uniaxial strain varying from −5% to +5% were calculated. The results are summarized in Figure S9 and S10. Via changing the magnitude of applied strains, the VBM and CBM levels of spin-up and spin-down electrons gradually shift. The trend is depicted in Figure 2b. Applying a compressive strain causes the spin-down VBM shifting upward and the CBM shifting downward, narrowing the band gap. At $\varepsilon_y \sim -2\%$, the VBM crosses the Fermi level, transforming the 1T'-CrS$_2$ into a spin-down half metal. As for the spin-up electron, the VBM shifts downward and thus the band gap opens up. In contrast, as the applied tensile strain increasing, the spin-up VBM moves upward and CBM level slightly moves downward. Under a strain $\varepsilon_y > 3\%$, the spin-up VBM and CBM cross the Fermi level and hence the spin-up band gap disappears. Meanwhile, the spin-down CBM level moves upward and the spin-down band gap increases. A tensile strain can transform the 1T'-CrS$_2$ to a spin-up half metal. The physical origins of observed electron level shifts will be discussed later.

In Figure 3, we summarize the band gaps of the spin-up and spin-down electrons as a function of uniaxial strain in *y*-direction. There are three zones corresponding to different electromagnetic properties of 1T'-CrS$_2$. In the blue zone, where a compressive strain $\varepsilon_y < -2\%$, the strained 1T'-CrS$_2$ is a spin-down half metal. In the green zone where strain is between −2% and 3%, the 1T'-CrS$_2$ remains as a FM semiconductor. When the tensile strain is higher than 3% (the red zone), it becomes a spin-up half metal. Using strain engineering to transform a 2D materials among FM semiconductor, spin-up half metal and spin-down half metal has not been reported before. A previous computational study showed that using a tensile strain of 10% could transform 1T-CrSe$_2$ from a FM metal to a spin-up half metal.[31] But no spin-down metal property was reported. Via inspecting Figure 1, the 2H-VSe$_2$ could be another possible case analogous to 1T'-CrS$_2$ (strain free state as FM semiconductor). Figure S11 shows the calculated the electronic DOS of 2H-VSe$_2$ under ±6% strain. VSe$_2$ remains as a semiconductor with little CBM and VBM shift under strains. Therefore, CrS$_2$ is the only TMD material with such tunable diverse electromagnetic properties that are highly desirable for spintronic applications. Note that the required strains for property transition, –2% and 3%, have been achieved in experiments for different types of 2D materials.[36,37]

To investigate physical mechanism of the strain induced half-metal transition, we decompose the spin-up and spin-down electronic structures of 1T'-CrS$_2$ under $\varepsilon_y$ = –6%, 0%, and +6% in Figure S12. Our previous analysis shows that the VBM levels continuously shift under strains and eventually lead to the property transition (Figure 2, S9, and S10). Understanding the relation between crystal deformation under external strain and the VBM levels should be the key. Figure S12 indicates that the VBM levels are mainly contributed from the S atoms. Indeed, the VBM charge density of the spin-up and spin-down electrons in Figure 4a confirm the VBM electrons are distributed around the S atoms. The spin-up VBM charge density (mainly around the S2 and S3 atoms) is distributed along the Cr-S bonds across the crystal layer. In contrast, the spin-down VBM charge density (around all S atoms) is distributed parallel to the crystal plane. Under a tensile strain, the interatomic distances between S2 and S3 atoms (one on top atomic layer and the other in bottom atomic layer) are reduced significantly (Figure 4b). As depicted in Figure 4c, the spin-up VBM electron clouds on S2 and S3 atoms would move closer, resulting in stronger interactions. The charge density distribution of the 1T'-CrS$_2$ under +6% strain in Figure 4a suggests a repelling interaction nature. As a result, the spin-up VBM level moves upward, consistently to our band structure analysis (Figure 2b). Note that a tensile strain increases the interatomic distances between S atoms along the crystal plane direction (i.e., $d_{12}$ and $d_{13}$). Thus, the spin-down VBM electron clouds move away from each other (Figure 4c). This could be the reason why the spin-down VBM level shows little shift in Figure 2b. Under a sufficiently large tensile strain ($\varepsilon_y$ > 3%), the spin-up VBM level will cross the Fermi level, transforming 1T'-CrS$_2$ to a spin-up half metal.

Similar understanding can be applied for the compressive strain induced spin-down half metal transition. In Figure 4b, a compressive strain reduces the interatomic distances between S atoms along the crystal plane direction (i.e., $d_{12}$ and $d_{13}$). The spin-down VBM electron clouds, which distributed along the crystal plane, move closer (Figure 4c). The VBM electron charge distribution profile of the 1T'-CrS$_2$ under −6% strain suggests a repelling interaction. As such, the VBM energy level of the spin-down electron shift upward, which is consistent to Figure 2b. Note that a compressive strain increases the S2-S3 distance. The spin-up VBM electron clouds move away from each other. There is little shift of spin-up VBM energy level, which is also consistent to Figure 2b. Under a sufficiently large compressive strain ($\varepsilon_y < -2\%$), the spin-down VBM level will cross the Fermi level, transforming 1T'-CrS$_2$ to a spin-down half metal. Overall, the interplay between the distinctive distribution of the spin-up and spin-down VBM electron clouds and crystal deformation under external strain is the reason for the strain induced half-metal transition.

**Phase stability and transition**

This section will discuss stability of the three phases of CrS$_2$, which is essential in engineering applications. The relative total energy of 2H, 1T' and 1T phases as a function of uniaxial strain along *y*-direction is shown in Figure 5a. Here the lattice constants of 1T' phase are taken as references to calculate the strain. The 2H phase is the most stable phase in the strain range from −8% to −1.5%. As $\varepsilon_y > -1.5\%$, the 1T' has a lower total energy than the 2H phase. In addition, the 1T phase has a lower total energy than the 1T' phase given $\varepsilon_y < -3.4\%$. In 2014, Reed group proposed the concept of strain induced phase transition in several TMD materials based on the energetic order changes of different phases under mechanical strains.[38] This concept was proved effective when strain is applied to trigger 2H-1T' phase transition in MoTe$_2$ under room temperature.[39] Similar strategies can be employed for our CrS$_2$ case.

From thermodynamics perspective, the common tangent lines for the energy vs. strain curves in Figure 5a determines the critical stress (under the stress loading condition), at which the 2H-1T' or 2H-1T phase transition would take place. Figure 5b demonstrates the phase transitions in terms of stress-strain relationships. The starting point of plateau at −6.5% represents the onset of the phase transition from 2H to 1T'. The plateau region from −6.5% to 4.5% represents the co-existence of the two phases. Beyond 4.5%, the 2H-1T' phase transition finishes with the 2H completely transition to 1T's phase. As for the 2H-1T phase transition, the phase transition starts at strain of −4.5% and ends at strain of 4.9%. In experiments, if we started from the most stable phase 2H, stretching the 2H phase about 2.5% and 4.5% would trigger the phase transition. The significant advances on strain engineering 2D materials in past few years have demonstrated that such strain magnitude is achievable in experiments.[40-43] In principle, applying a compressive strain could transform the 1T' phase to 1T. Figure S13a shows the result. The phase transition starts at −5.8% and ends at −1.1%. It is also possible to apply such compressive strain in experiments on 2D TMD material.[40,44] But applying compressive strain on 2D materials is more difficult due to possible mechanical buckling issue.

In our DFT calculations, spontaneous transition was not observed when stretching or compressing $CrS_2$ within the strain range shown in Figure 5a and 5b. This suggests that all the three phases are metastable under these strain conditions. Some energy barriers must be overcome and the phases transition kinetics in experiments should be considered. Then the climbing image nudged elastic band (CI-NEB) method was employed to calculate the transition energy barrier. Figure 5c shows the results for the 2H-1T' phase transition under a strain from −6% to 4%. With increasing stretching 2H phase, the energy barrier quickly reduced from ~0.8 eV (at −6%) down to only 0.25 eV per formula unit at a strain of 4%. The phase transition could take place under appropriate experimental condition, following the analysis in Ref. 36, in which the energy barrier from 2H to 1T' phase is about 0.88 eV/f.u. for $MoTe_2$ and phase transition was realized in experiment.[39]

Our DFT calculations results for the transition energy barrier between 1T' and 1T phase under different strains (0%, −2%, −4% and −6%) are shown in Figure S13b. Be aware there were technical issues to carry out the CI-NEB calculations between a AFM phase and a FM phase. Thus, we adopted two-step methods to semi-quantitatively estimate the energy barrier (details in Supplementary Information). The estimated energy barrier is much lower than those in 2H-1T' phase transition, *i.e.*, 29.6meV/f.u. and 14.8meV/f.u. at −4% and −6% strain, respectively. It is more feasible to use strain to induce phase transition between 1T-AFM and 1T'-FM.

In addition to strain, there are plenty of experimental and theoretical studies that demonstrate other methods to induce the phase transitions in TMD materials. For example, using DFT calculations, Li *et al* found that electron doping could induce 2H to 1T/1T' phase transition in MoTe$_2$.[45] Later Wang *et al* experimentally confirmed the predicted transition.[46] Light or electron beam irradiation could also induce phase transition.[47,48] For example, Kang *et al* deposited Au nanoparticles on monolayer 2H-MoS$_2$ and triggered 2H-1T phase transition through light irradiation. The light irradiation generated hot electrons, which were then doped into MoS$_2$ and trigger the phase transition. Some other methods related to charge doping could also induce phase transition in TMDs, such as transition metal atoms alloying, lithium or other foreign atoms absorption and vacancy.[49-54] In addition, annealing, microwave assistance, and pressure could also be used to trigger phase transition in TMDs.[55,56]

**Novel strain-controlled spintronics device designs**

The diverse electron/magnetic properties of CrS$_2$, particularly the strain induced spin-up an and spin-down half metals, could offer a promising possibility to integrate different functionalities on a single piece of CrS$_2$ nanosheet by purposely control local phase structures or strain conditions for spintronic devices. Spintronics is an emerging technology brings great opportunities for many novel devices with spin-dependent effects, such as magnetic field sensors, read heads for hard drives, galvanic isolators, and magnetoresistive random access memory (MRAM).[1,2,9,57,58] The core of these devices is a spin valve. The conventional basic structure of a spin valve has three layers: two FM layers sandwiched one NM interlayer. In light of its 100% spin polarization ratio, half-metal is the most attractive candidate for the two FM layers.[59,60] The spin valve can be either GMR or TMR, if the interlayer is conductor or insulator.[1,58] The resistance is low (conductive state) when the magnetization (spin polarization) in ferromagnetic layers are parallel and is high (non-conductive state) when they are antiparallel. There is some difference between GMR and TMR spin valves. The role of metallic interlayer in GMR is conducting electrons and screening interaction between two side FM layers. The semiconductor or insulator interlayer in TMR spin valve is to provide tunneling barrier instead of conducting electrons. Spin polarized electrons could transport across the interlayer due to the tunneling effect in TMR. Thus, TMR based spin valve is also called magnetic tunnel junction (MTJ).

The diverse properties of $CrS_2$ can satisfy the requirements of both GMR and TMR spin valves. A spin-valve device can be fabricated in a single piece of $CrS_2$ nanosheet as a lateral heterostructure. In this heterostructure, the 1T' phase (under strain) serves as the two FM layers and the interlayer component can be either 1T-$CrS_2$ (metallic) or 2H-$CrS_2$ (semiconductor) phase (making a GMR or TMR spin valve). In practice, an array of spin-valves (multi-channels) would be fabricated for a nonvolatility, infinite endurance and fast random access logic devices, *e.g.*, MRAM as a promising candidate for 'universal memory'.[58] A simple design of strain-controlled two-channels or bits (denoted as I and II) spin valve logic device is demonstrated in Figure 6. Due to the unique property of $CrS_2$, *i.e.*, positive/negative strain induced spin-up/spin-down half metals, our $CrS_2$ spin valve device can be operated using strain instead of external magnetic fields, as illustrated in Figure 6. When applying the same strains on the two FM layers, both of them will be either spin-up or spin-down half metal. As a result, only the spin-up or spin-down electrons could be conducted through, which can be termed as digit '1' or '−1'. If different types of strains were applied, one FM layer would be spin-up half metal but the other one would be the spin-down half metal. There will be no electric current, *i.e.*, the off state, termed as digit '0'. By purposely applying strains on the two FM layers (the 1T'-$CrS_2$ phase), different information would be the output, *e.g.*, the nine ($3^2$) possible information in Figure 6 for the two-channel device. By increasing the channel number *n*, we can get much more information than the conventional charge-based devices, $3^n$ vs. $2^n$. Figure S14 depicts the CBM and VBM levels of 1T'-$CrS_2$ under uniaxial strain compared with 1T and 2H phase. The CBM of 2H is always higher than 1T' CBM meaning 2H phase could effective separated two half-metal layers as an tunneling layer in TMR. In addition, the CBM/VBM of 1T phase is always lower than that of 1T' phase leading electrons effectively conducted through 1T in GMR. Thus, the band alignment of $CrS_2$ different phase could satisfy basic requirements both in GMR and TMR.

There is also an opportunity to design a magnetic field controlled TMR spin valve integrated on a single piece $CrS_2$ nanosheet. The device structure is shown in Figure S15. The basic structure is similar with our strain controlled TMR spin valve (Figure. 6), but the main difference is that magnetic field controlled TMR spin valve needs one more 1T-$CrS_2$ AFM layer as a pinning layer. The pinning layer is used to fix the magnetic moment of the neighbor FM layer through exchange effect between FM and AFM. Thus, the structure of this device is 1T'(half-metal)/2H(semiconductor)/1T'(half-metal)/1T(AFM-metal). The aim of these layers is generating spin-polarized electrons under external magnetic field, providing tunneling barrier to separate to half-metal layers, providing reference spin polarization, and pining FM layers.

The Curie temperature ($T_C$) of FM semiconductor or half metal is an important parameter for spintronics devices.[61] It is highly desirable to have $T_C$ higher than room temperature. We adopted Heisenberg model to estimate the $T_c$ of 1T'-CrS$_2$ as reported by previous studies.[31,62] The details of our method are shown in Supplementary Information. In Figure S16, as strain changes from −6% to 6% along *y*-direction, the 1T'-CrS$_2$ $T_C$ varies between ∼ 377.3K to ∼ 1358.1K. This is not a surprise, since previous works indicate that Cr compound often have a higher $T_C$, such as CrTe$_2$ and CrC. [31,35,61]

To confirm the feasibility of integrating different phase on one single piece CrS$_2$ nanosheet for above spintronics devices, 2H-1T' and 1T-1T' heterostructures both along *x* and *y* direction are constructed in Figure S17. After fully relaxed in our DFT calculations, all these lateral heterostructures are stable. In the past few years, there are significant advances of experimental techniques to fabricate 2D lateral heterostructure. For example, the MoTe$_2$/MoS$_2$ and WS$_2$/MoS$_2$ lateral heterostructures were fabricated obtained by using chemical vapor deposition (CVD).[22,63,64] Besides, 2H/1T' MoTe$_2$ and MoS$_2$ lateral heterostructure were fabricated using CVD or chemical exfoliation methods.[23,25] There is a good opportunity to fabricate the designed CrS$_2$ lateral heterostructures (Figure 6) in experiments. As suggested in previous studies, using strain to control the magnetization alignment consumed less energy than the conventional methods.[20,21] In addition, the simple structure (less interfaces) and atomically sharp interfaces of our 2D CrS$_2$ lateral heterostructure could reduce electron transport resistance. Therefore, CrS$_2$ could be an excellent candidate for spintronics and our findings provide inspiration for designing a low-power consumption, simple structure and multifunction spintronics devices.

## Conclusion

In this paper, using DFT calculations, we systematically surveyed all IV, V and VI group TMDs and identified CrS$_2$ as a unique material suitable for spintronic device designs. It has the most diverse electronic and magnetic properties that meet the requirements of spintronics. It has a non-magnetic (direct band gap) semiconductor 2H phase and an AFM metal 1T phase. Its 1T' phase shows FM ground state and is a (indirect band gap) semiconductor with 0.26 eV and 1.92 eV for spin-up and spin-down electrons, respectively. The most interesting observation is that a uniaxial tensile or compressive strain along *y*-direction can switch 1T' phase into a spin-up or down half metallic state. This unique feature is particularly useful for spintronic device designs. As an example, a prototype spin valve device is proposed based on 2D $CrS_2$ lateral heterostructure. This device is operated under strain instead of external magnetic fields. The new features of such 2D lateral heterostructure device, such as strain control, simple structure, and atomically sharp interface, could significantly reduce energy consumption, which remains as a challenge in miniaturizing spintronic devices. Our results indicate that $CrS_2$ is an excellent material candidate to develop simple structure, low power consumption, and multifunction spintronics nanoscale device

**Computation method**

Our DFT calculations were conducted by using the Vienna Ab-inito Simulation Package (VASP).[65] The spin polarized Perdew−Burke−Ernzerhof (PBE) exchange-correlation functional[66] and projector augmented wave method[67] were adopted. Considering PBE calculations will underestimate the band gap, the electronic structures are also calculated by using the Heyd–Scuseria–Ernzerhof (HSE06) hybrid functional.[68] The cutoff energy was set to 400 eV, while the Monkhorst−Pack k-point mesh[69] was set to 21 × 19 × 1 for 1T'-$MX_2$ primitive unit cell and 1×$\sqrt{3}$×1 supercell of 2H and 1T phase. Atomic positions and lattice constants were fully relaxed until total energy difference and forces were less than $10^{-4}$ eV and 0.005 eV/Å, respectively. A higher convergence criterion for total energy, $10^{-6}$ eV, was adopted to perform self-consistent calculation calculate and obtain electronic structures. Periodic boundary condition was applied in all three directions. A vacuum space of at least 20 Å was applied along *z*-directions to avoid the interaction between two periodic images. The value of the in-plane uniaxial strain along *y*-direction is defined as $\varepsilon_y = (b - b_0)/b_0 \times 100\%$, where *b* and $b_0$ are the in-plane lattice constants of the strained and unstrained 1T' monolayers, respectively. Climbing image nudged elastic band (CI-NEB) calculations[70] were performed to estimate phase transition barriers of 2H-1T' and 1T-1T' transition. The maximum residual forces of NEB calculations are smaller than 0.01eV/Å. Considering NEB calculation cannot describe magnetic order reversion process of 1T (AFM)-1T'(FM) transition, we first calculated nonmagnetic 1T-1T' transition process to approximate reaction coordinates. Then self-consistent calculation was performed on each coordinate with AFM and FM magnetic order. The magnetic order with a lower total energy at each coordinate is adopted to approximately describe the energy barrier between 1T-AFM and 1T'-FM phase.

**Acknowledgement**

The authors gratefully acknowledge the support of NSFC (Grants Nos. 11974269, 51728203, 51621063, 51601140, 51701149, 51671155, 91963111), the support by 111 project 2.0 (Grant No. BP2018008), the National Science Basic Research Plan in the Shaanxi Province of China (2018JM5168) and Innovation Capability Support Program of Shaanxi (Nos. 2018PT-28, 2017KTPT-04). J.D. also thanks the Fundamental Research Funds for the Central Universities. J.Z.L. acknowledges the support from ARC discovery projects (DP180101744) and HPC from National Computational Infrastructure from Australia. This work is also supported by State Key Laboratory for Mechanical Behavior of Materials and HPC platform of Xi'an Jiaotong University. The authors would also like to thank Mr. F. Yang and Dr. X. D. Zhang at Network Information Center of Xi'an Jiaotong University for support of HPC platform.

**Author contributions statement**

J.D., J.Z.L. and S.Y. designed the simulations and the framework of this research. K.C. carried out the first-principles calculations. Y.Y. carried out the phonon spectrum calculation and other detailed calculations. K.C., J.D. and J.Z.L. wrote the paper. Q.S. and T.C. interpreted the data and help to draw figures. J.S. and X.D. performed some data analysis and provide many suggestions.

**Additional information**

**Competing financial interests:** The authors declare no competing financial interests.

# Figure 1

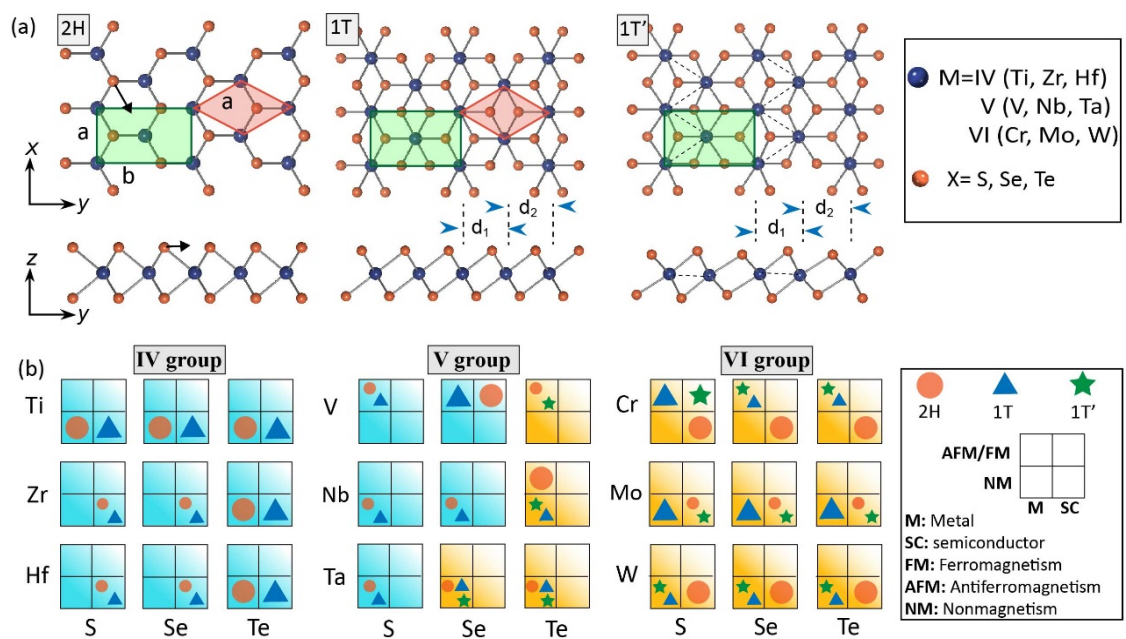

**Figure 1 Crystal structures and physical properties of 2H, 1T and 1T' phase of group IV, V and VI TMDs.**
**(a)** Crystal structure of 2H, 1T and 1T'-$CrS_2$. The primitive unit of 2H and 1T are highlighted in orange rhombus. The primitive unit of 1T' corresponds to the 1x$\sqrt{3}$ supercell of 2H and 1T that is highlighted in the green rectangle. The 2H phase can be derived via collective top layer lateral slide to the center of hexagon denoted by the black arrow in the top view and side view. The 1T' phase can be derived via the structure distortion of 1T phase. These features are general in all TMDs monolayers. **(b)** Electronic and magnetic properties of IV, V and VI TMDs. Note that $CrS_2$ has the most diverse properties, including antiferromagnetic metal, ferromagnetic semiconductor and non-magnetic semiconductor corresponding to 1T, 1T' and 2H phase, respectively.

| Table 1 Lattice constants, magnetic order and relative energy of 2H, 1T and 1T' phase of $CrS_2$. | | | | | | | | |
|---|---|---|---|---|---|---|---|---|
| **Phase** | **Structure information** | | | | magnetism | ΔE (eV/f.u.) | **Band gap (eV)** | |
| | $a$(Å) | $b$(Å) | $d_1$ | $d_2$ | | | Spin up | Spin down |
| **2H** | 3.04 | 5.26 | | | NM | 0 | 1.35 | |
| **1T** | 3.28 | 5.4 | 3.16 | 3.16 | AFM | 0.3734 | 0 | |
| **1T'** | 3.32 | 5.62 | 2.99 | 3.55 | FM | 0.3436 | 0.26 | 1.92 |

**Figure 2**

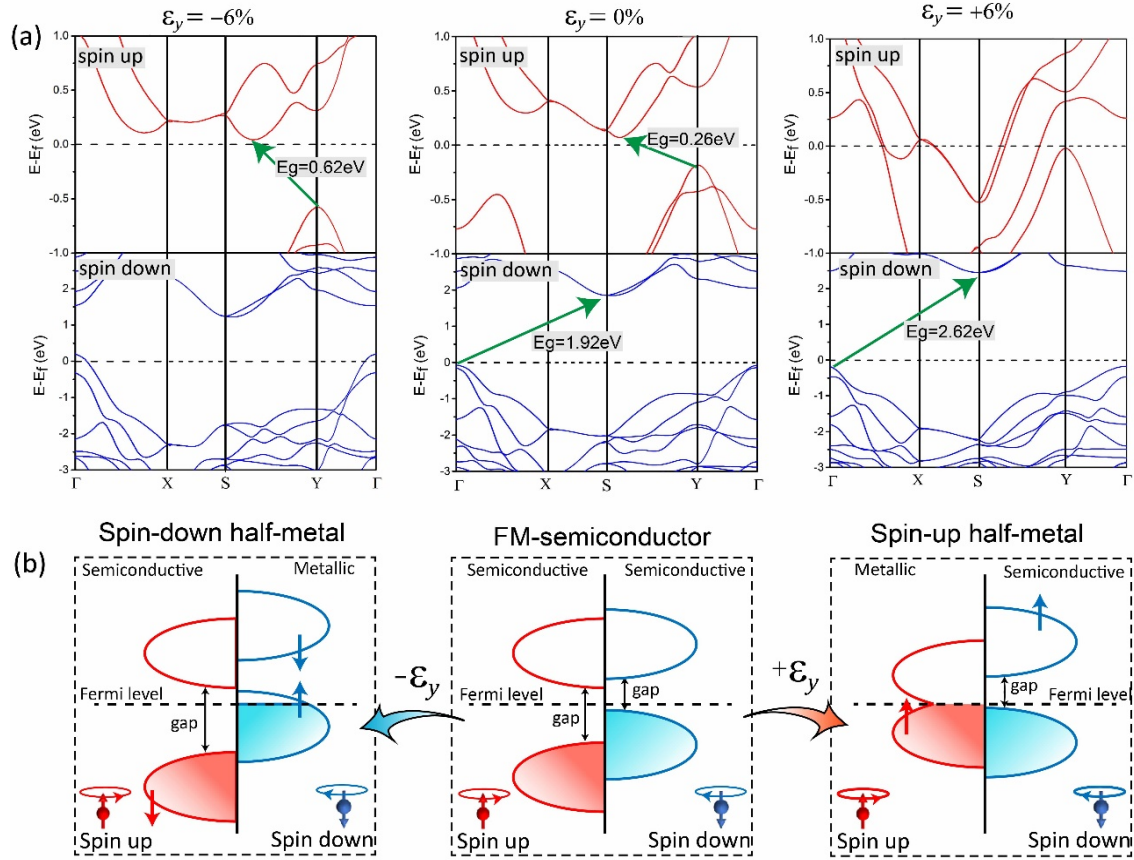

**Figure 2 Electronic band structures of 1T'-CrS$_2$ under different strains along *y*-direction. (a)** spin polarized band structure of 1T' phase under −6%, 0% and +6% strain along *y*-direction calculated by using HSE06. The spin-up band gap disappears under +6% strain while the spin-down gap disappears under −6% strain, corresponding to spin-up half-metallic and spin-down half-metallic states, respectively. The band structures at other strain are summarized in supplementary information Figure S9 and S10. **(b)** A diagram to illustrate electron structure changes under tensile (+$\varepsilon_y$) and compressive (−$\varepsilon_y$) strain. The tensile strain causes spin-up VBM and CBM shifting upward and downward, respectively. The CBM will cross Fermi level, transforming 1T'-CrS$_2$ to spin-up half metal. The compressive strain causes spin-down VBM shifting upward and cross Fermi level leading 1T'-CrS$_2$ transits to spin-down half-metal.

**Figure 3**

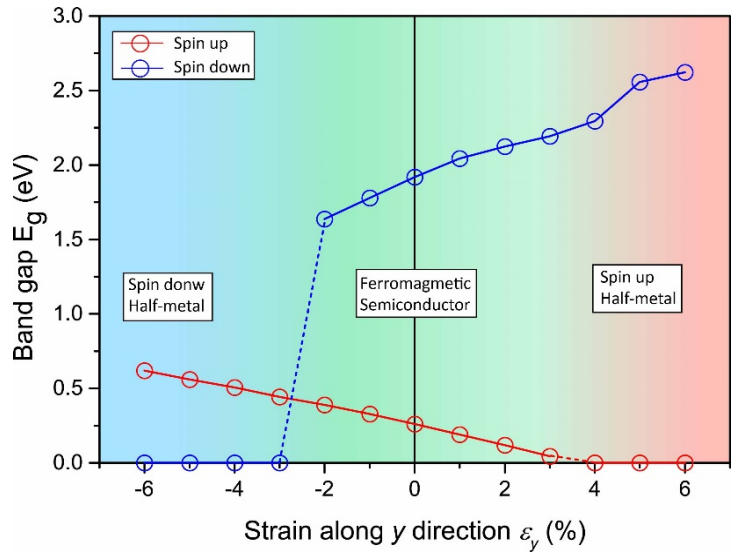

**Figure 3 Strain engineering to transform 1T'-CrS$_2$ from FM semiconductor to spin-up half-metal or spin-down half-metal states.** Band gap $E_g$ of spin-up and spin-down electrons as a function of mechanical strain in $y$-direction. At strain free state, 1T'-CrS$_2$ in a FM semiconductor with spin-up $E_g \sim 0.26$eV and spin-down $E_g \sim 1.92$eV. Applying a positive strain $\varepsilon_y$ leads to reduction of spin-up $E_g$ and increase of spin-down $E_g$. At $\varepsilon_y \sim +4\%$, the spin-up $E_g$ diminishes, *i.e.* a spin-up half-metal state. Appling a $\varepsilon_y$ negative leads to an opposite trend. At $\varepsilon_y \sim -3\%$, spin-down $E_g$ closes, *i.e.* spin-down half-metal.

**Figure 4**

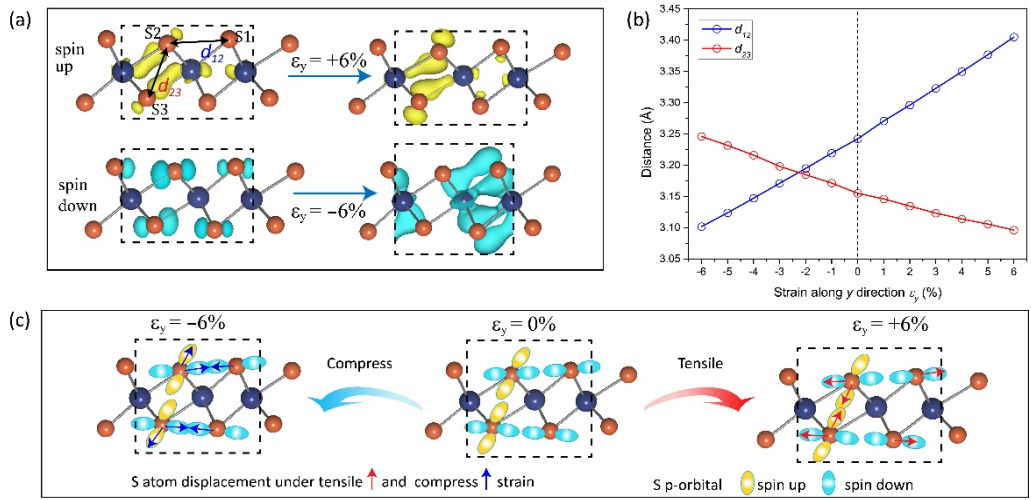

**Figure 4 The crystal and electronic structures analysis to reveal the mechanism of strain induced spin-up and spin-down half-metal switching of 1T'-CrS$_2$.** (a) The electron charge density of spin-up and spin-down VBM of CrS$_2$ changes under ±6% strain (isosurface value is 10$^{-3}$). Both spin-up and spin-down VBM electrons mainly concentrate at S atoms. The difference is that spin-up electrons along Cr-S bond direction but spin-down electrons along *y*-direction in *y-z* plane. (b) The distance of S1-S2 ($d_{12}$) and S2-S3 ($d_{23}$) changing with *y*-direction strain from −6% to +6%. (c) The diagram of tensile and compressive strain induced VBM spin polarized electrons changing. Under tensile strain, spin-up VBM electrons on S2 and S3 tend to overlap due to $d_{23}$ decreasing making spin-up electrons cross Fermi level and turn to metal. However, spin-down electrons will not be affected because of the increasing $d_{12}$. Under compressive strain, this tendency is reversed. Under +6% strain, spin-up electrons mainly concentrate between non-dimerization part. Under −6% strain, spin-down electrons mainly concentrate between dimerization part.

**Figure 5**

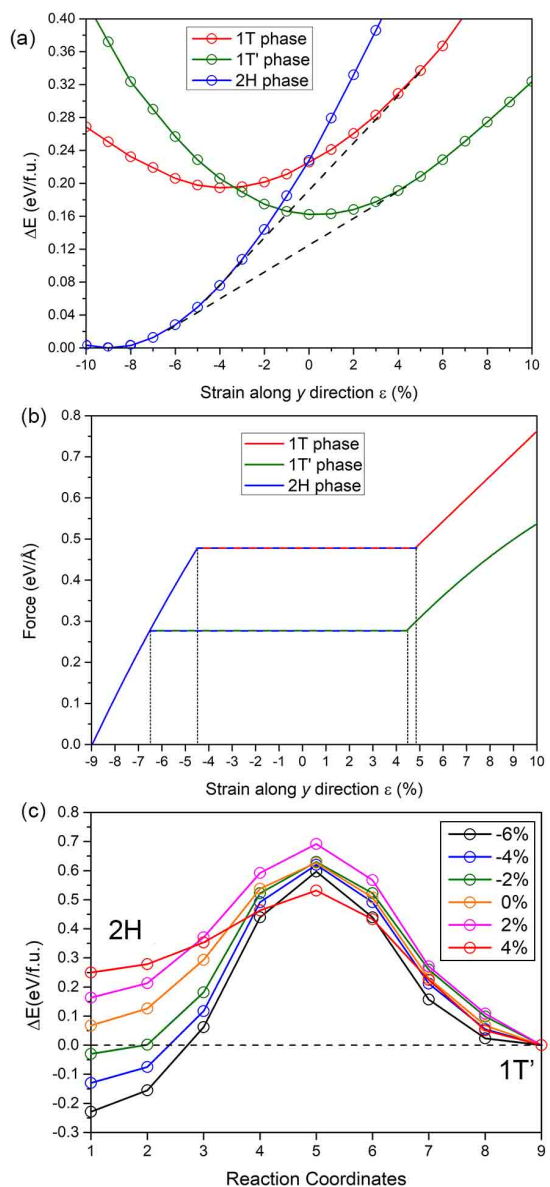

**Figure 5 Phase stability and transition between 1T, 1T' and 2H-CrS$_2$. (a)** The relative energy of 2H, 1T and 1T' phases as a function of uniaxial strain along $y$-direction. The strain is defined with the equilibrium state of 1T' phase. The dashed lines are the common tangents. **(b)** 2H-1T' and 2H-1T phase transition process respect to (a). In step 1, 2H phase deforms elastically without phase transition. Beyond some critical strain (−6.5% for 1T' and −4.9% for 1T) in step 2, the lowest free-energy path is a common tangent manifesting a coexistence regime where both phases coexist in mechanical equilibrium. At the strain level of step 3, the lowest energy phase is composed of 100% 1T' at $\varepsilon_y$ = 4.5% and 100% 1T at $\varepsilon_y$ = 4.9%, completing the mechanically induced phase transition. **(c)** NEB calculation results for the phase transformation between 2H and 1T' under $\varepsilon_y$ = −6%, −4%, −2%, 0%, 2% and 4% strain. The results show substantial energy barriers to separate 2H and 1T' phase.

**Figure 6**

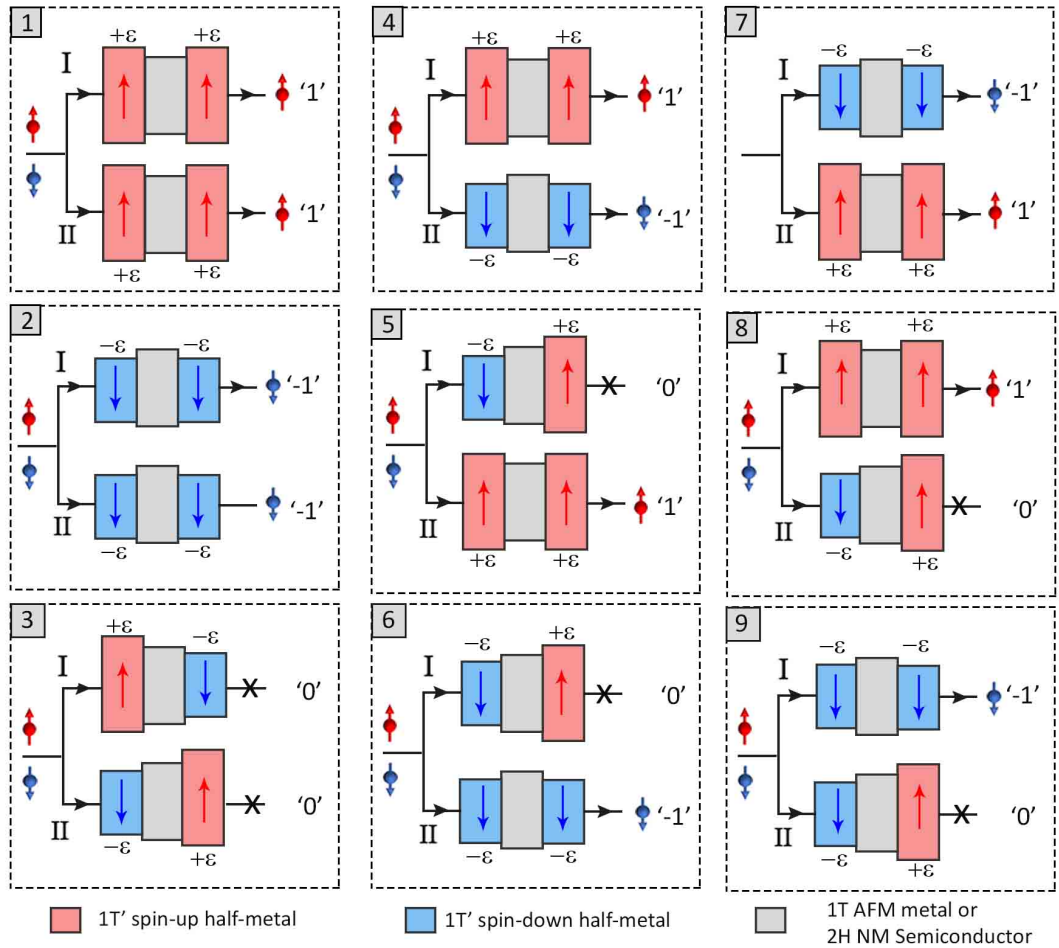

**Figure 6. A diagram for a strain-controlled spin valve logic device with two channels and each channel being based on a single CrS$_2$ nanosheet.** Strain can be used to switch spin up or down half-metal and control local structure transition. There could exits two types heterostructure for spin valve depending on the interlayer phase structure, 1T'/1T/1T' for GMR and 1T'/2H/1T' for TMR. The spin polarized current can be defined as signal '1' and '−1' for spin- up and down electrons, respectively. Once two side half-metal are set as spin up and down half-metal respectively, both spin up and down electrons cannot be conducted which could be regarded as OFF state, defined as signal '0'.

# Diverse electronic and magnetic properties of CrS$_2$ enabling novel strain-controlled 2D lateral heterostructure spintronic devices

Kaiyun Chen[1], Junkai Deng[1,*], Yuan Yan[1,2], Qian Shi[1], Tieyan Chang[1], Xiangdong Ding[1], Jun Sun[1], Sen Yang[1,*] and Jefferson Zhe Liu[2,*]

[1] MOE Key Laboratory for Nonequilibrium Synthesis and Modulation of Condensed Matter, State Key Laboratory for Mechanical Behavior of Materials, Xi'an Jiaotong University, Xi'an 710049, China

[2] Department of Mechanical Engineering, The University of Melbourne, Parkville, VIC 3010, Australia

*junkai.deng@mail.xjtu.edu.cn; *yangsen@mail.xjtu.edu.cn; *zhe.liu@unimelb.edu.au

*Corresponding authors

**Phase stability and properties for IV, V and VI group transition metal dichalcogenides (TMDs).** 1T, 1T' and 2H phase of all possible IV, V and VI group TMDs are calculated and shown in Figure S1. 1T' phase cannot be observed in IV group TMDs and 1T is the ground state other than 2H phase. In V group, 2H phase is always the ground state and 1T' phase begins to appear which is a metastable phase. VI group is similar to V group but 1T' phase can exit in all TMDs. The metallic, semiconductivity and magnetism are concluded in Figure S2. The magnetism can only be observed in V/CrX$_2$. Obviously, CrS$_2$ possess the most various properties instead of other TMDs. 1T-CrS$_2$ is antiferromagnetic metal and 2H-CrS$_2$ is nonmagnetic semiconductor according to Figure S3.

**Phonon spectrum of single-layer 1T'-CrS$_2$.** Despite 1T' is not the ground state of CrS$_2$, the phonon spectrum of 1T'-CrS$_2$ (Figure S4) can also inform us that this phase can relative stable exits.

**Magnetic order and stability of CrS$_2$.** Various magnetic orders of 1T and 1T' phase CrS$_2$ are calculated as depicted in Figure S5. The stability of these order is shown in Table S1 in which 2H is set as a reference. 1T phase show stripe AFM order which energy is higher than all possible magnetic order similar to previous results.[1,2] The ground state of 1T' phase is FM magnetic order which is also a metastable phase. The spin polarized charge density of 1T and 1T' phase is shown in Figure S6. The corresponding density of state (DOS) is shown in Figure S7. It can be observed that the spin polarization on the S atom of 1T' phase is stronger than that of 1T. This fact informs Cr-Cr interaction will change from direct exchange interaction in 1T phase to superexchange interaction in 1T' phase.

**Band structures from under different strain conditions.** The band structure of 1T'-CrS$_2$ under 0%, +6% and −6% strain is also calculated by using PBE method in Figure S8. Compared with HSE06 results in Figure 2, PBE band structures can also describe the half-metal switching process under different strain conditions. The main difference is HSE06 calculation can give us more accurate band gap value. The band structures from HSE06 other strain from −5% to +5% are shown in Figure S9 and S10.

**Decomposed band structures.** Despite HSE06 is good to describe band gap value, it is hard to calculate the charge density on a special band and k point due to its self-consistent band structure calculation. Considering the PBE can also describe the half-metal switching under different strain, the decomposed band structures are calculated in Figure S12. For spin-up band structures, it can be observed that the electrons on S2 and S3 band (denoted by blue) along Γ-X and T-Γ will cross the Fermi level after $\varepsilon_y = +6\%$ is applied leading semiconductive 1T' phase turn to spin-up half-metal. For spin-down band structures, the electrons on S1, S2 and S3 band at Γ will cross Fermi level under $\varepsilon_y = -6\%$ and then leading 1T' phase turn to spin-down half-metal.

**Strain controlled 1T'-1T phase transition.** NEB method is hard to calculate the energy barrier between two phases with different magnetic order because only one magnetic order can be defined in one NEB process. To estimate the phase transition energy barrier between 1T-AFM and 1T'-FM, the nonmagnetic transformation pathway was firstly calculated by NEB method. The aim of this step is to estimate an approximate phase transition pathway. Second, self-consistent calculation was performed at each coordinate in this pathway by using FM and AFM magnetic order respectively. The magnetic order with lower total energy at each coordinate is adopted shown in Figure S13b.

**Band alignment of $CrS_2$.** Figure S14 is the band alignment of $CrS_2$. The CBM of 2H is always higher than VBM of every phase, which means the barrier between 2H and 1T' is enough to prevent electron directly conduction. Besides, the VBM/CBM of 1T phase is lower than that of 1T' phase, which means electrons are allowed transported through 1T in GMR devices.

**Magnetic field controlled TMR spin valve.** The magnetic field controlled TMR spin valve structure is similar to the strain-controlled spin valve, the most different is that magnetic field TMR has an AFM layer. Due to the exchange interaction between FM and AFM layer, FM magnetic moment could be pinned to avoid rotation under magnetic field. But, another FM can be freely rotated under an external magnetic field. The interlayer is 2H tunnelling layer providing the tunnelling barrier and separating to FM layers. The corresponding structure is shown in Figure S15. The structure of this device is FM-free layer/tunnelling layer/FM layer/AFM pinning layer and the phase in this device is 1T'/2H/1T'/1T.

**Curie temperature ($T_C$) estimation.** The $T_C$ of 1T'-CrS$_2$ are estimated according to the Heisenberg model reported by previous studies[1,3]:

$$H = -J \sum_{<i,j>} S_i \cdot S_j \quad (1)$$

For CrS$_2$ monolayer, $J$ can be derived as:

$$J = \frac{E_{AFM} - E_{FM}}{N|S|^2} \quad (2)$$

where |S| is 1, N is the number of magnetic atoms in one unit cell, and $E_{FM}$ and $E_{AFM}$ are the energies of FM and AFM state of 1T' phase. Then $T_C$ could be estimated as:

$$T_C = \frac{2}{3 \cdot K_B} J \quad (3)$$

The $T_C$ of 1T'-CrS$_2$ under different strain is shown in Figure S16. From −6% to +6%, $T_C$ changes from 377.3K to 1358.1K. All $T_C$ are higher than room temperature, which informs the ferromagnetism of 1T'-CrS$_2$ is stable and could be useful in room temperature.

**Lateral heterostructures of CrS$_2$.** Different phase lateral heterostructures are calculated shown in Figure S17. The combination junction of distinct two phases, 1T'/1T, 1T/2H and 1T'/2H can stably exist. It provides the possibility for CrS$_2$ different phase heterostructure junction which is the basis for electronic or spintronic devices.

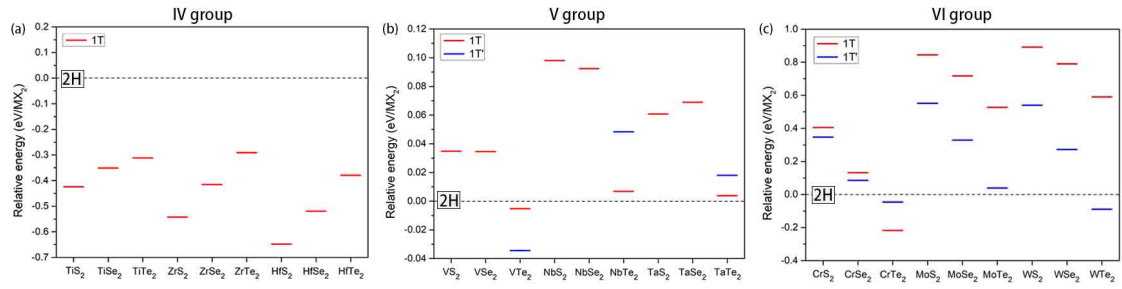

**Figure S1 The stability relationship of 2H, 1T and 1T' phase in IV, V and VI TMDs.** (a) IV TMDs. (b) V TMDs. (c) VI TMDs. The energy of 2H phase is set as reference shown by black dot line.

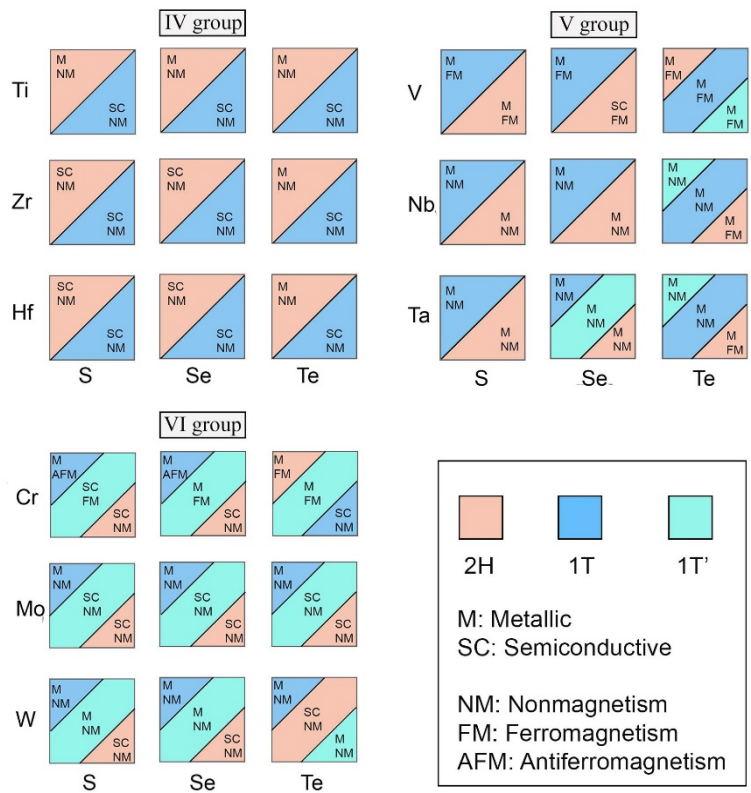

**Figure S2 All structures and their corresponding properties of IV, V and VI TMDs.** It can be clearly seen that $CrS_2$ has the most various properties, antiferromagnetic metal, ferromagnetic semiconductor and non-magnetic semiconductor corresponding to 1T, 1T' and 2H phase.

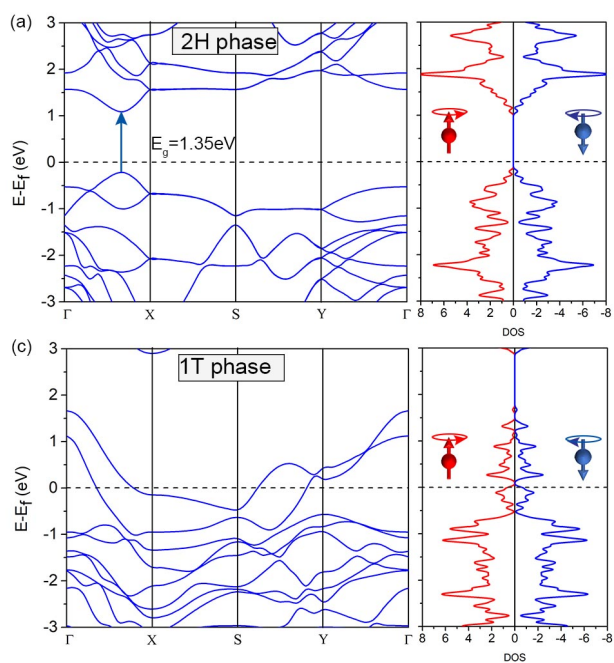

**Figure S3 Electronic structures of CrS$_2$.** **(a)** Band structure and DOS of Non-magnetism 2H phase. **(b)** Band structure and DOS of anti-ferromagnetism 1T phase calculated with HSE06 functional.

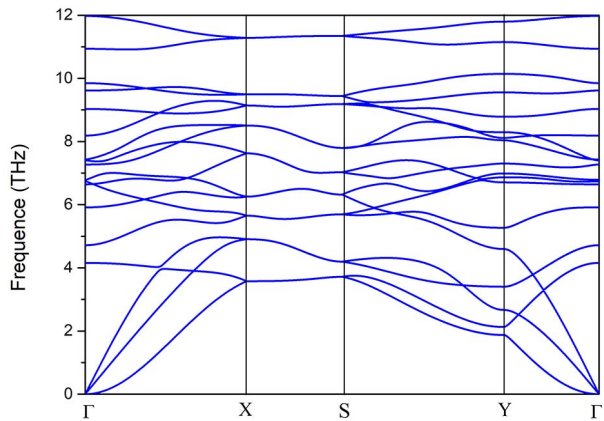

**Figure S4** Phonon spectrum of single-layer 1T'-CrS$_2$.

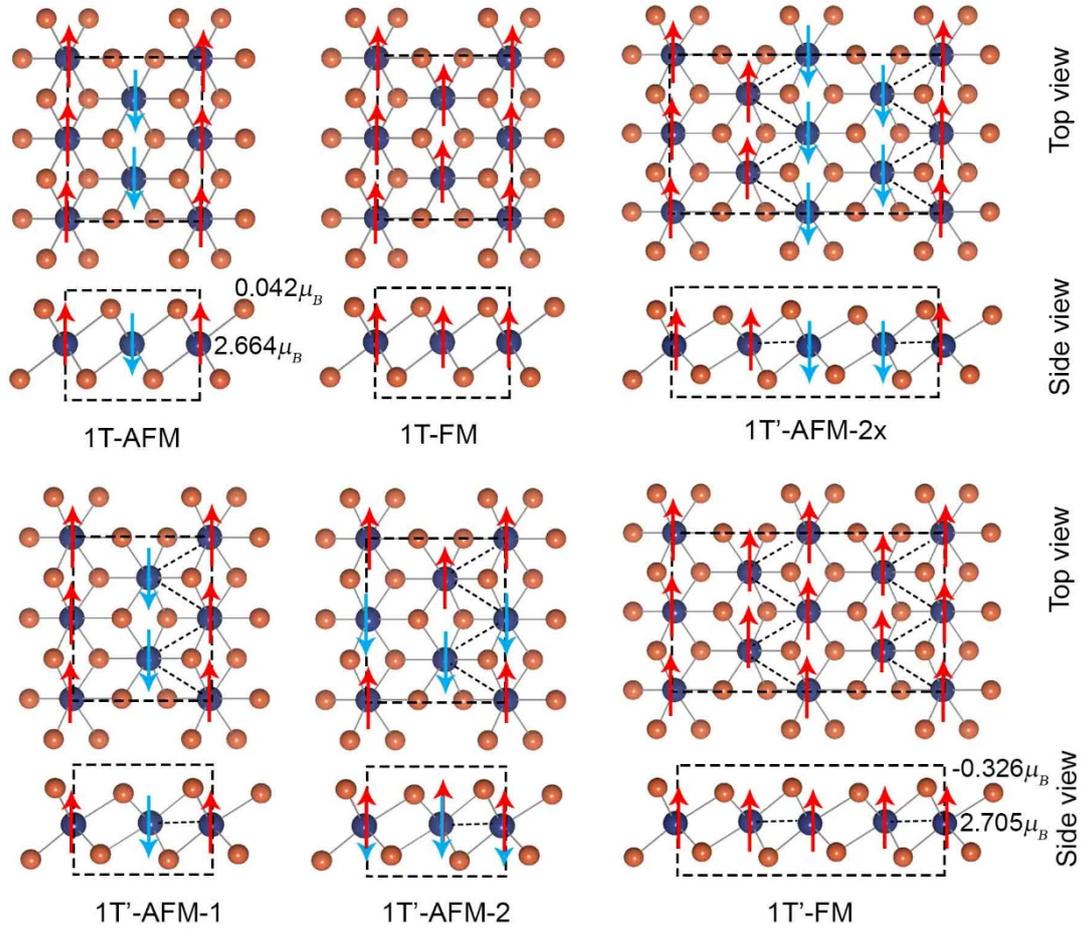

**Figure S5 Ferromagnetic and antiferromagnetic order of 1T and 1T' phase.** The red and blue arrows represent spin up and down, respectively.

**Table S1.** Relative energy of 2H, 1T and 1T' phase with different magnetic order. 2H NM phase is used as a reference.

| $CrS_2$ | 1T | | 1T' | | | | 2H |
|---|---|---|---|---|---|---|---|
| Magnetic order | FM | AFM | FM | AFM-1 | AFM-2 | AFM(2x) | NM |
| E (eV/f.u.) | 0.441 | 0.378 | 0.347 | 0.496 | 0.434 | 0.4841 | 0 |

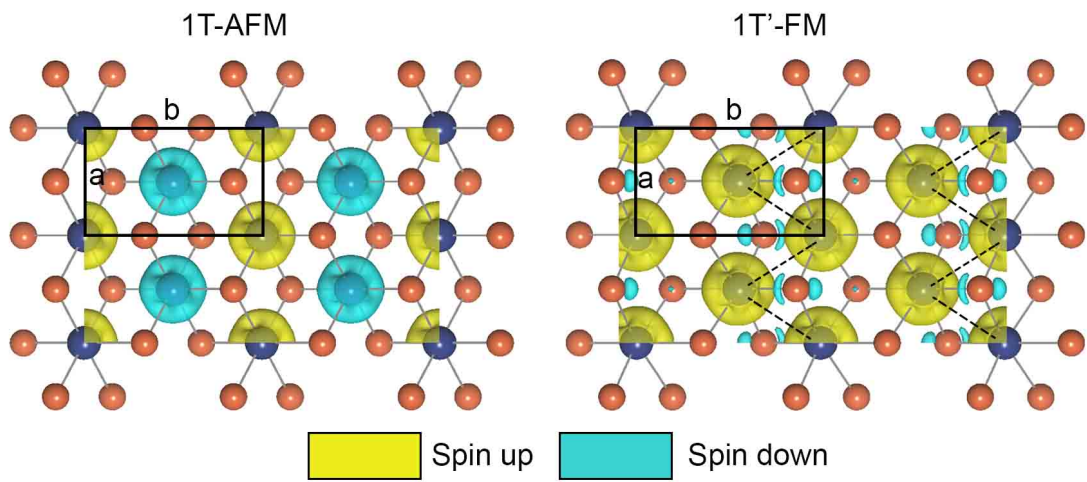

**Figure S6. Spin polarized charge density of CrS$_2$.** 1T phase is antiferromagnetic and 1T' is ferromagnetic. The isosurfurace is 0.01 eV.

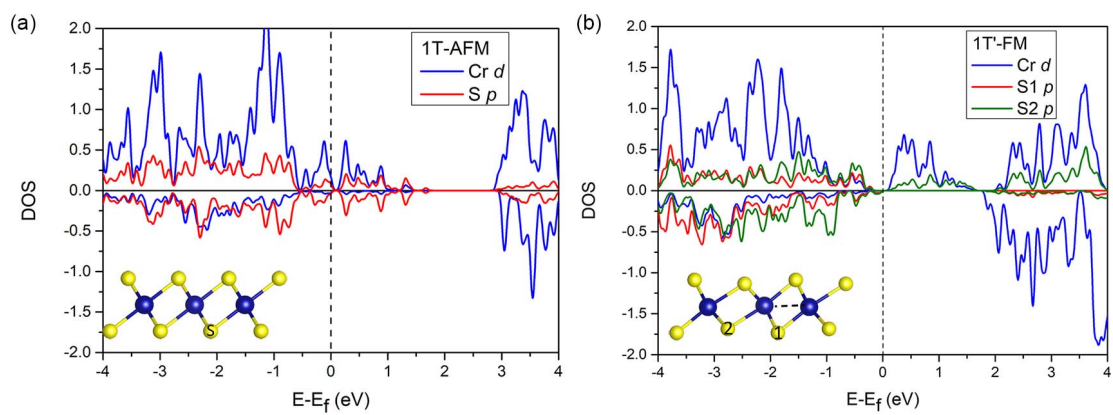

**Figure S7**. Cr *d*-orbital and S *p*-orbital DOS of (a) 1T-AFM-CrS$_2$ and (b) 1T'-FM-CrS$_2$ calculated with the HSE06 functional.

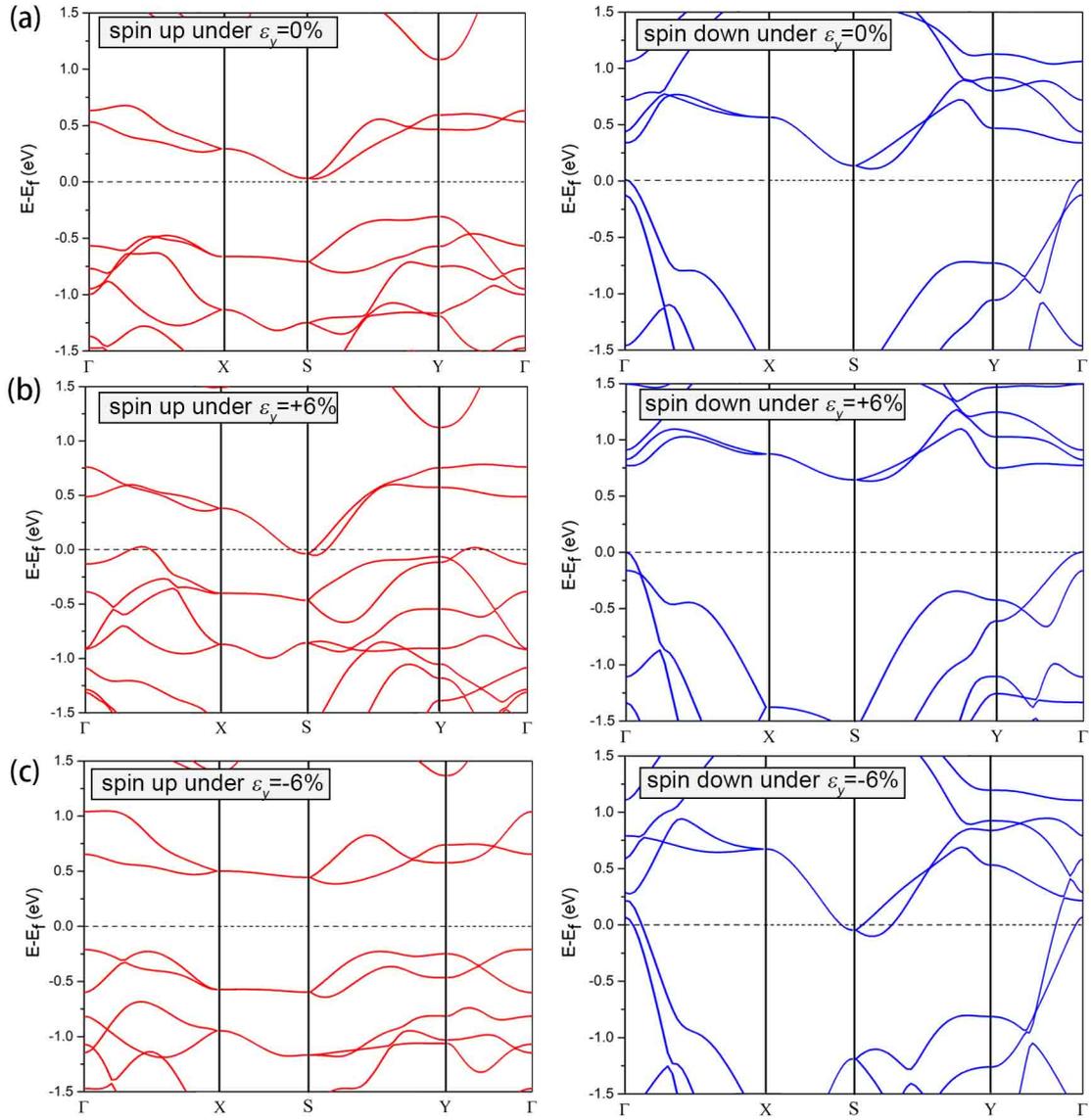

**Figure S8 Electronic band structures of 1T'-CrS$_2$ (FM) calculated with PBE functional.** (a) Spin-up (left) and down (right) band structure of 1T' phase without strain applied. (b) Spin-up (left) and down (right) band structure of 1T' phase under +6% strain along *y*-direction. (c) Spin-up (left) and down (right) band structure of 1T' phase under −6% strain along y direction. The spin-up band gap disappears under +6% strain while the spin-down gap disappears under −6% strain, corresponding to spin-up half-metallic and spin-down half-metallic, respectively.

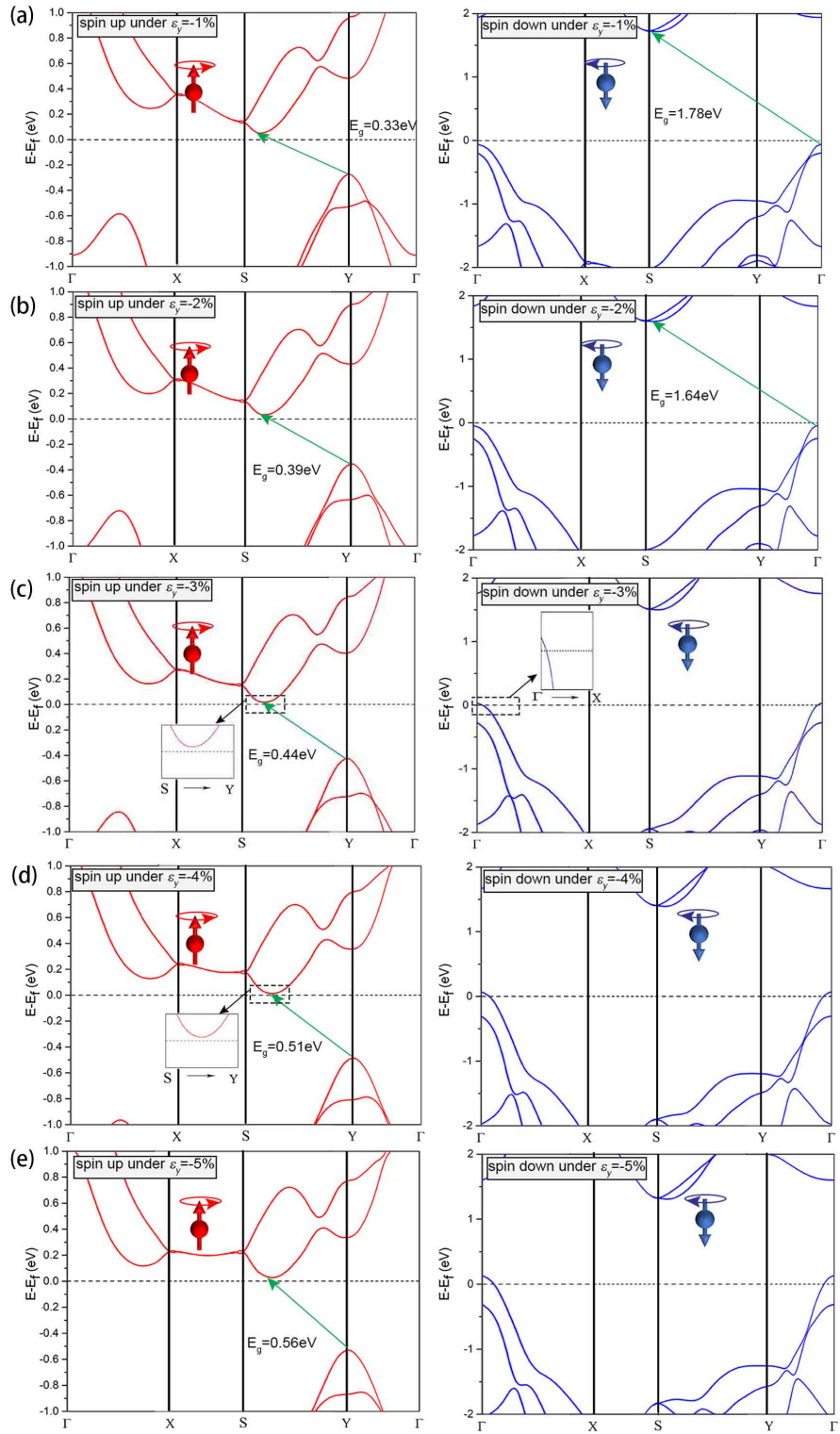

**Figure S9 Spin polarization band structure of 1T'-CrS$_2$ under different compressive strain along *y* direction.** **(a)** −1% strain. **(b)** −2% strain. **(c)** −3% strain. **(d)** −4% strain. **(e)** −5% strain. The left and right figure corresponds to spin-up and down band structure, respectively. The spin-down gap disappears when compressive strain beyond −2% and 1T'-CrS$_2$ (FM semiconductor) transfers to spin-down half-metal.

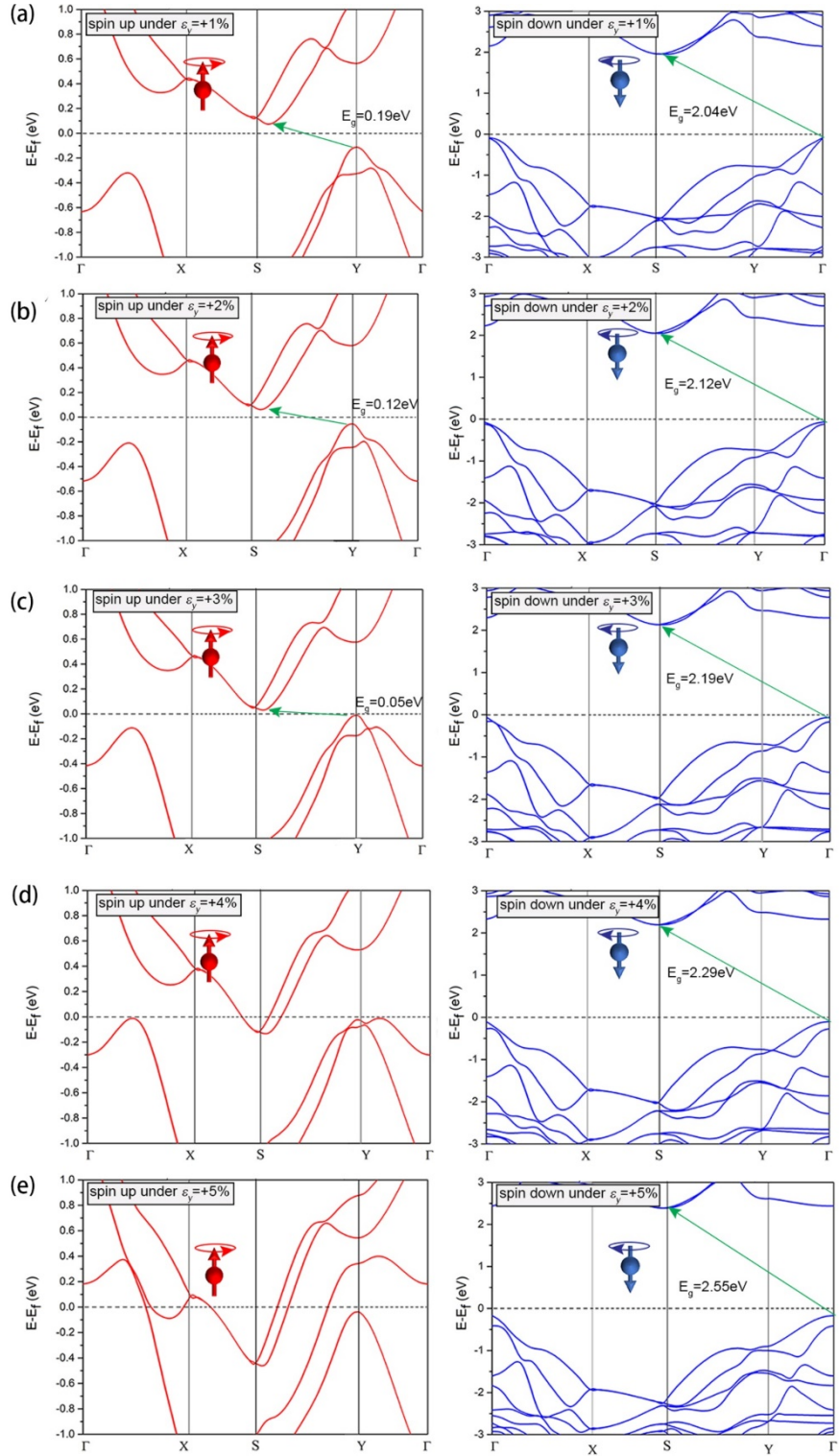

**Figure S10 Spin polarization band structure of 1T'-CrS$_2$ under different tensile strain along *y* direction. (a)** +1% strain. **(b)** +2% strain. **(c)** +3% strain. **(d)** +4% strain. **(e)** +5% strain. The left and right figure corresponds to spin-up and down band structure, respectively. The spin-up gap disappears when strain beyond 3% and 1T'-CrS$_2$ (FM semiconductor) transfers to spin-up half-metal.

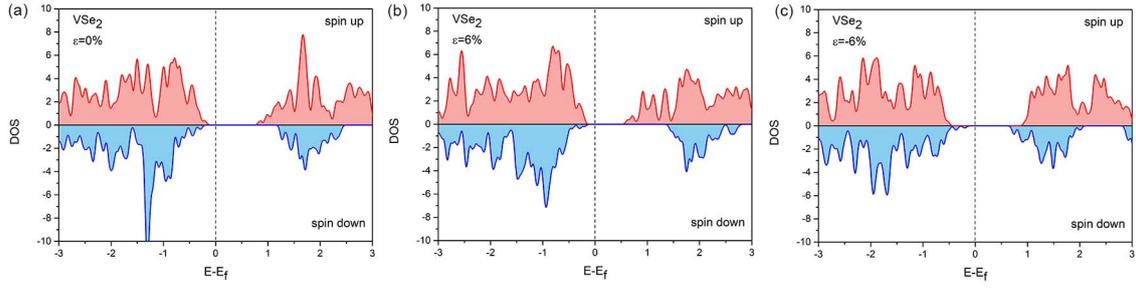

**Figure S11 The density of state of 2H-VSe$_2$ under different uniaxial strain along *y* direction calculated by HSE06.** (a) 0%. (b) +6%. (c) −6%. VSe$_2$ is always ferromagnetic semiconductor.

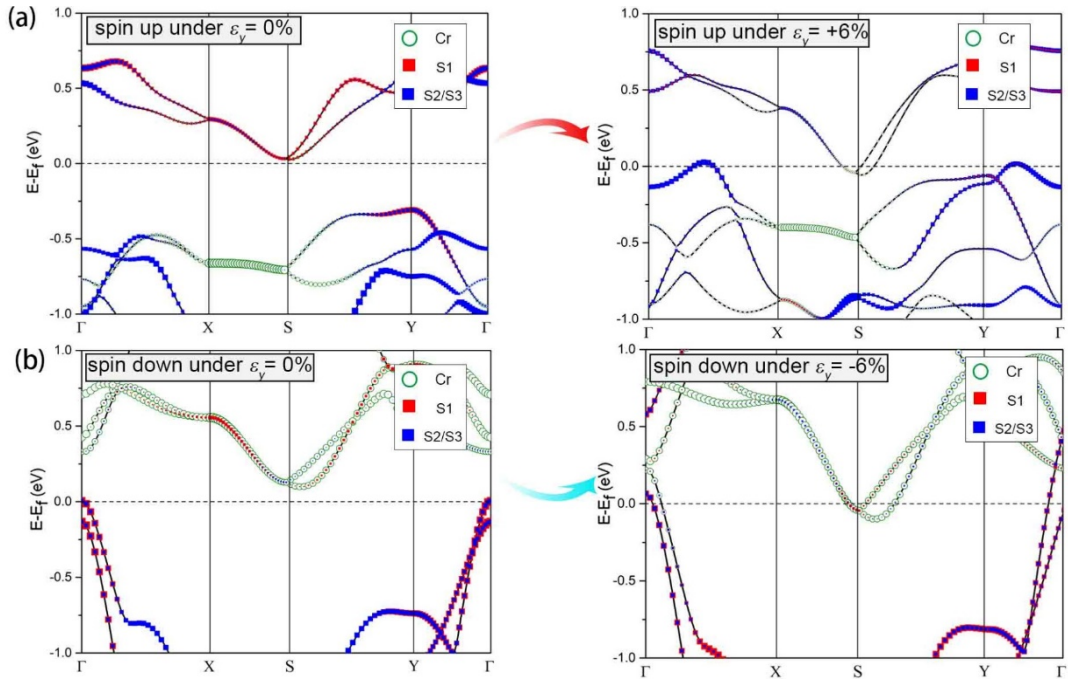

**Figure S12 The decomposed spin polarized band structure of 1T'-CrS$_2$ under +6% and −6%, respectively.** Under +6% strain, the spin up band gap disappears due to S2 band beyond Fermi level. Meanwhile, the spin down band gap disappears due to both S1 and S2 beyond Fermi surface under −6% compress. The number of S atom corresponds to Figure 4a.

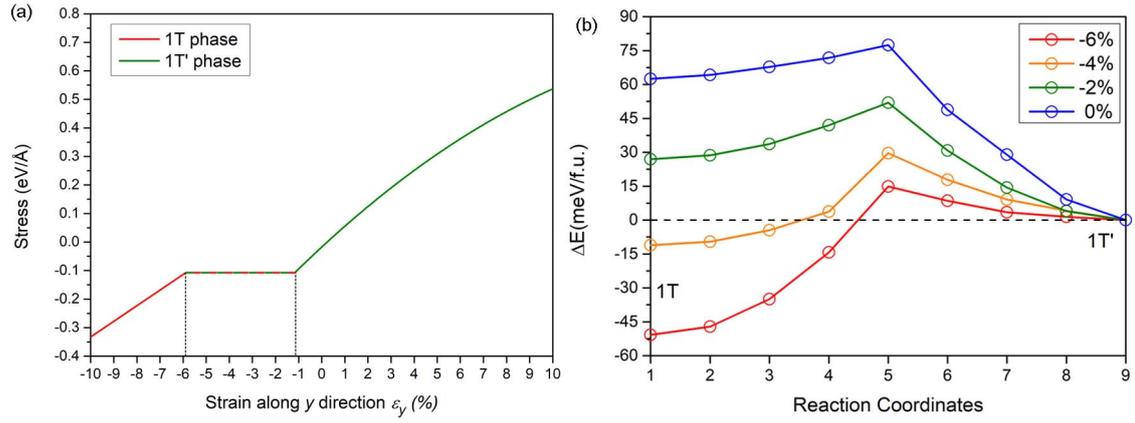

**Figure S13 Strain controlled phase transition between 1T' and 1T phase.** (a) 1T'-1T phase transition process respect to Figure 5a. (b) **(d)** NEB calculation results for the phase transformation between 1T and 1T' phase under $\varepsilon_y$= −6%, −4%, −2% and 0%.

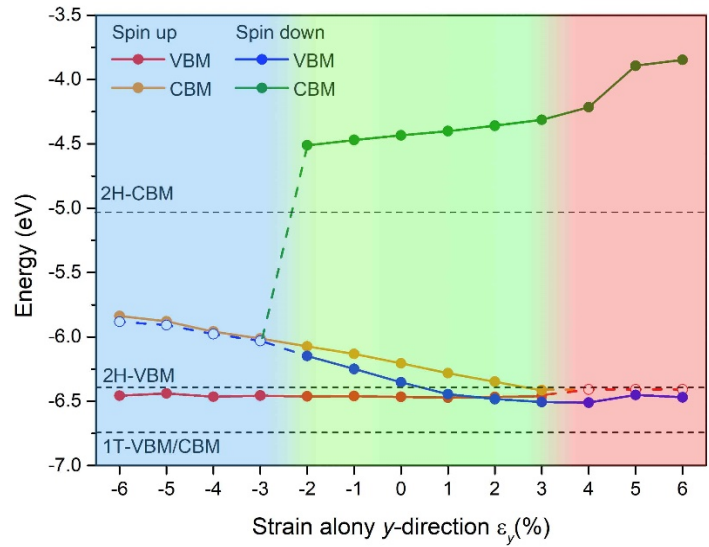

**Figure S14** CBM and VBM of 1T'-CrS$_2$ under uniaxial strain along *y*-direction compared with 1T and 2H phase.

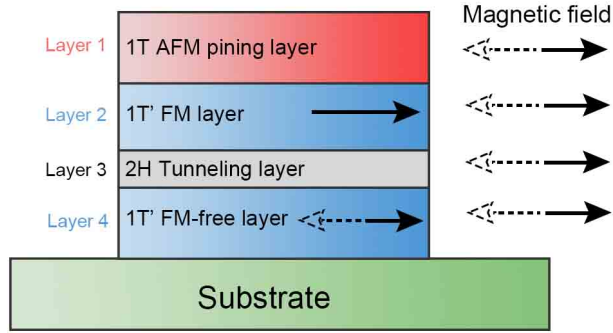

**Figure S15** Magnetic field controlled TMR spin valve.

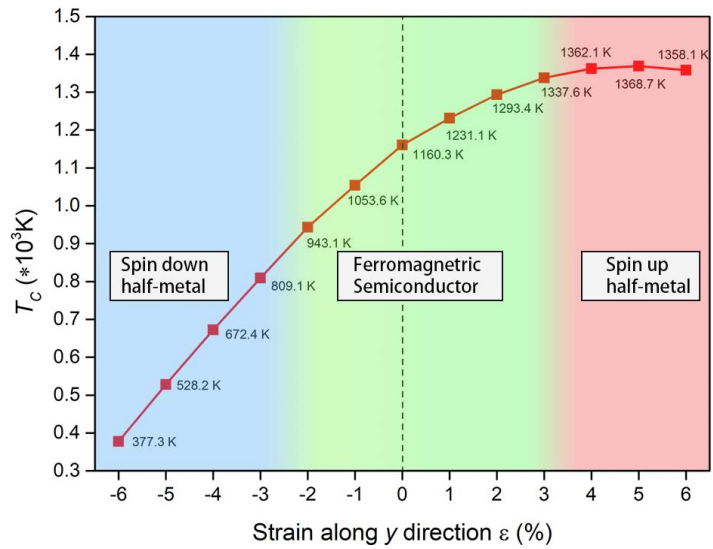

**Figure S16 Curie temperature (T$_C$) of 1T'-CrS$_2$ under *y* direction strain.** The semiconductor T$_C$ is from 943.1K to 1337.6K as strain changes from −2% to 3%. It means that ferromagnetic order could relatively exits under room temperature.

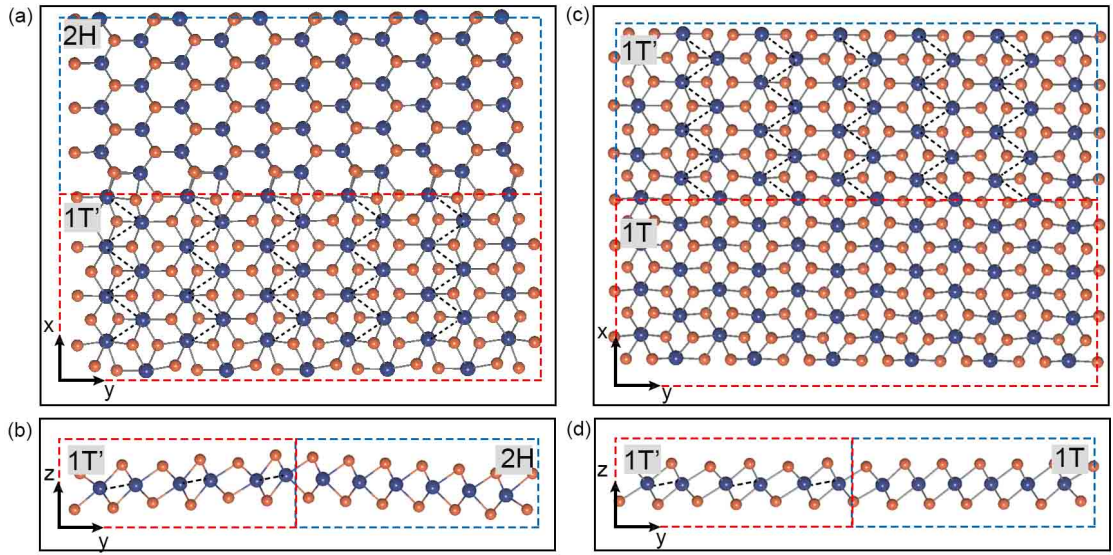

**Figure S17 Different phase heterostructure junctions of CrS$_2$.** Heterojunction structure between 2H and 1T' along (a) *x*-direction and (b) *y*-direction. Heterojunction structure between 1T and 1T' along (c) *x*-direction and (d) *y*-direction. The combination junction of distinct two phase, 1T'/1T and 1T'/2H can stably exits.